\documentclass[11pt,a4paper]{article}
\usepackage[utf8]{inputenc}
\usepackage[english]{babel}
\usepackage{amsmath}
\usepackage{amsfonts}
\usepackage{amssymb}
\usepackage{graphicx}
\usepackage{cite}
\usepackage{hyperref}
\usepackage{lmodern}
\usepackage[left=2cm,right=2cm,top=2cm,bottom=2cm]{geometry}
\author{Subhankar Roy\footnote{subhankar@gauhati.ac.in(corresponding author)} \& Sagar Tirtha Goswami\footnote{sagartirtha@gauhati.ac.in}\\{\small Department of Physics, Gauhati University, India}}
\date{}
\title{\textbf{Leptons and other forces of nature}}

\begin{document}
\maketitle
\begin{abstract}
Assuming that neutrinos are not Majorana fermions and the right handed Dirac neutrino does not exist, we propose a model in which the second and the third generations of the leptons are composites,  while the first generation is fundamental.  The composite states are formed by the fundamental leptons and two new fundamental hidden scalar particles.  In addition,  there exist two hidden forces besides the SM interactions.  The gauge symmetry $SU(2)_L\otimes U(1)_{Y}\otimes U(1)_h\otimes SU(2)_h$ of the electroweak and the hidden forces breaks down to $U(1)_{Y}\otimes U(1)_h$ after the spontaneous symmetry breaking (SSB). We explain the neutrino masses in terms of the binding dynamics of a hidden force.  The phenomenon of neutrino oscillation can also be explained by our model in a dynamical framework of the hidden forces
\end{abstract}

\section{\label{Section 1}{Introduction}}

 Understanding nature at the ``fundamental'' level is the quest of particle physics. The quest is, however, an eternal one. The nature reveals itself only to the extent of `energy' spent on the quest to know it, which, in particle physics, translates to the collision energy of the colliders. The notion of what is the fundamental level of nature is therefore not conclusive,  but can be thought of as the current limit of our understanding of nature.

So far, what we know about ``fundamental'' particles and their interactions is encoded in the standard model (SM) of particle physics \, \cite{Glashow:1961tr,Weinberg:1967tq,Salam:1968rm,Glashow:1970gm}. According to the SM,  there are twelve fundamental fermions: six leptons ($e, \mu, \tau, \nu_e, \nu_\mu, \nu_\tau$) and six quarks ($u,d,c,s,t,b$). Their interactions are mediated by twelve bosons: the photon ($\gamma$), the carrier of electromagnetic interaction; the three gauge bosons ($W^{\pm},  Z^0$), the carriers of the weak interaction,  and the eight gluons ($g$),  the carriers of the strong interactions. The fermions are categorised into three generations or families as follows.\\

Leptons:
\begin{eqnarray}
\begin{array}{ccc}
\begin{pmatrix}
e\\
\nu_e
\end{pmatrix} &\begin{pmatrix}
\mu\\
e_\mu
\end{pmatrix} & \begin{pmatrix}
\tau\\
\nu_\tau
\end{pmatrix}\\
1^{\text{st}} & 2^{\text{nd}} & 3^{\text{rd}}
\end{array} 
\end{eqnarray}

Quarks:
\begin{eqnarray}
\begin{array}{ccc}
\begin{pmatrix}
u\\
d
\end{pmatrix} &\begin{pmatrix}
c\\
s
\end{pmatrix} & \begin{pmatrix}
t\\
b
\end{pmatrix}\\
1^{\text{st}} & 2^{\text{nd}} & 3^{\text{rd}}
\end{array} 
\end{eqnarray}

Besides,  there exists a spin-zero particle,  named  Higgs boson ($H$),  which is associated with the origin of masses of the fundamental particles,except neutrinos.

The symmetry group of the SM is considered as $SU(3)_c \otimes SU(2)_L \otimes U (1)_Y $. The $SU(3)_c$ is associated with the strong interaction. The Glashow-Weinberg-Salam (GWS) model relates the unified form of the electromagnetic (EM) and the weak forces,  known as the electroweak (EW) force with the gauge group $SU(2)_L \otimes U(1)_Y$.  After the Higgs mechanism\,\cite{Englert:1964et, Higgs:1964ia, Higgs:1964pj},  the symmetry of the Lagrangian breaks down spontaneously and the fundamental particles except the neutrinos obtain masses.  Along with the quarks and leptons, the gauge bosons $W^{\pm}$ and $Z^0$ also acquire masses.  The $\gamma$,  however,  remains massless after the spontaneous symmetry breaking.  Hence,  even though the gauge symmetry is broken, still the $U(1)_{Q}$ symmetry is preserved. The success of the GWS model is confirmed by the discovery of the $W^{\pm}$\,\cite{Arnison:1983rp, Banner:1983jy} and $Z^0$\,\cite{Arnison:1983mk, Bagnaia:1983zx} bosons, and finally the Higgs boson\,\cite{Aad:2012tfa,Chatrchyan:2012ufa}.   

The neutrinos within the standard model remain massless, because the GWS model does not allow right handed neutrinos.  But the results of the neutrino oscillation experiments\, \cite{Davis:1968cp,SNO:2001kpb,Super-Kamiokande:2001ljr,SNO:2002tuh,Bionta:1987qt,Super-Kamiokande:1998kpq,KamLAND:2002uet,DayaBay:2012fng,K2K:2002icj,T2K:2011ypd,T2K:2013ppw,Kamiokande-II:1989hkh}, if analysed in the light of Pontecorvo's proposition\, \cite{Pontecorvo:1967fh} require that neutrinos have mass.  Experimentally,  it is not possible,  so far, to identify the masses of the individual neutrinos exactly,  but information of the upper bounds on their mass scales or on the sum of their masses can be obtained from various experimental data, e.g., from various decay experiments like  the $\beta$ decay experiments \,\cite{Formaggio:2021nfz,Otten:2008zz,KATRIN:2019yun,ParticleDataGroup:2020ssz} and from the cosmological data \,\cite{Lesgourgues:2012uu,ParticleDataGroup:2020ssz} etc. . Although these bounds do not necessarily prove that neutrinos are massive,  they leave sufficient space for the possibility that neutrinos might have mass within the set limit. These considerations go against the prediction of the SM or the GWS model.  However,  the experiments so far,  have not found any evidence of the right handed neutrinos \,\cite{Goldhaber:1958nb,Boser:2019rta}.  So,  the GWS model might not be wrong in excluding them.  In order to solve this conjecture,  the theorists relate neutrinos with Majorana nature and this hypothesis takes us beyond the SM. The Majorana nature is associated with the indistinguishability of particle and antiparticle states for electrically neutral fermions like neutrinos \,\cite{Majorana:1937vz,Furry:1939qr}. One important evidence for the Majorana nature of neutrinos would be the observation of lepton number violating processes like neutrinoless double beta decay, which is not confirmed by the experiments, yet \,\cite{Dolinski:2019nrj}.  Again there is an issue with the smallness of the neutrino mass compared to the other leptons.  To explain this,  various models have been put forward,  e.g.,  seesaw models \,\cite{Gell-Mann:1979vob,Mohapatra:1979ia,Glashow:1979nm,Magg:1980ut,Foot:1988aq},  radiative models \,\cite{Wang:2016lve,Ma:2016mwh,Ma:2006km} etc. , but most of these models assume either the existence of right handed neutrinos (Dirac scheme) or that the neutrinos are Majorana particles (Majorana scheme) \,\cite{Cai:2017jrq}.  In addition to the problem of neutrino mass,  there are some interesting issues regarding the Pontecorvo-Maki-Nakagawa-Sakata (PMNS) paradigm\, \cite{Bilenky:1978nj,Maki:1962mu} of neutrino oscillations. It says that a neutrino flavour state is a linear superposition of the mass eigenstates. A mass eigenstate has a definite mass, a definite momentum and hence a definite energy.  The neutrino flavour oscillation occurs as a result of the differences in the time evolution of the mass eigenstates\,\cite{Bilenky:2013wna}.  The probability of conversion of one flavour into another flavour of neutrino is given by  \,\cite{1969PhLB...28..493G,Bilenky:1975tb,Bilenky:2013wna},  

\begin{equation}
P_{\nu_{l} \longrightarrow \nu_{l'}}\sim (sin^2{2\theta})  sin^2{ (\Delta m^2 L/4E}),
\end{equation}

where $\Delta m^2$ is the mass squared difference between two neutrino mass eigenstates,  $\theta$ is the mixing angle,  $E$ is the neutrino enregy and $L$ is the distance between the source and the detector.
It is interesting to note that unlike the quarks whose mass eigenstates are recognized by the strong interaction and the flavour eigenstates by the weak interaction\,\cite{Cabibbo:1963yz,Kobayashi:1973fv},  the neutrino mass eigenstates are not recognised by any known interaction.  Now,  $
P_{\nu_{l} \longrightarrow \nu_{l'}}$ is measured in the neutrino oscillation experiments.  So, we can see that $\Delta m^2$ and $\theta$ are rather the results obtained from the experiments than being theoretical predictions.  Hence,  one cannot conclude that the $\nu$ oscillation,  which is a physical phenomenon, is the outcome of a quantum interference effect. One cannot, therefore,  reject an alternative explanation for $\nu$ oscillation.  In addition,  the SM provides no explanation for another issue,  namely the dark matter (DM)\, \cite{Rubin:1970zza,Rubin:1980zd,Bertone:2004pz},  a kind of matter which does not interact with electromagnetic radiation. The DM contributes  $27\,\%$ to the total mass energy content of the universe,  whereas the visible matter contributes by an amount of $5\,\%$. 

Both of these problems related to neutrino and DM open new pathways to think beyond the SM.  But the  triumph of the SM is beyond doubt and hence the fundamental framework of the GWS model should be respected.  Rather,  one alternative strategy could be to review the primary axioms of the SM in the light of the present observations.  It is to be noted that,  regarding the mass of the neurino,  the problem lies in the Yukawa term of the Lagrangian,  which denotes the couplings of the fundamental particles with the Higgs field.  If the neutrino were not a fundamental particle,  the Yukawa term for the neutrino would hardly be needed.  The neutrinos could get masses through other mechanisms.  Perhaps the idea of `compositeness' could give answers to the origin of $\nu$ masses.

 In the context of the compositeness of the SM particles,  we note that there is a set of models,  known as `composite models' that address the structural issues\, \cite{Lyons:1982jb,Buchmuller:1985nn} with the SM by claiming that the particles in the SM are not fundamental, rather they are composed of new fundamental particles\, \cite{Fritzsch:1981zh,Harari:1980ez,Pati:1974yy,Dugne:1999ez}. These models are not focused enough on the mainstream particle physics research, especially after the discovery of the Higgs boson as these models do not consider the Higgs mechanism as a source of mass generation.  But since the Higgs mechanism cannot provide masses to the neutrinos, we should not disregard the importance of the composite models in this regard.

In the present work, we try to modify the GWS model without deviating much from the original philosophy.  If, for the sake of argument, it is confirmed in  future experiments that neither right handed Dirac neutrinos exist nor are neutrinos Majorana particles,  then explaining the origin of neutrino mass will become difficult in  terms of the contemporary models.  We will try to seek this answer by looking into the possible coexistence of hidden matter with SM particles and the presence of hidden forces of nature along with the EW interactions. To keep the discussion simple,  we shall consider the leptons only. 

The paper is organised in the following way.  In Section \ref{Section 2},  we present the axioms of this extended GWS model and elaborate the same.  In Section \ref{Section 3},  we attempt to explain some phenomena including the masses of the composite leptons and the neutrino oscillations based on our model.  And finally in Section \ref{Section 4},  we provide the summary and discussion of our proposed model.

\section{\label{Section 2}The Extended GWS Model}
We propose a composite model in which the first generation of leptons is elementary while the second and third generations are composites.  We assume that neutrinos are Dirac particles and there exists no right handed neutrino. The mass of the neutrino appears because of its composite nature.  But if neutrinos become composite particles, then the charged leptons cannot remain untouched by this fact. This particular choice of elementary and composite particles is also supported by the observation that $e, \mu$ and $\tau$ are identical in all properties except their masses and the fact that only the electron is a stable particle.  A similar choice was also adopted in\, \cite{Shaw:1980ku}.  
\vspace{3mm}

The axioms of our model are given below.

\begin{enumerate}
   \item There are two fundamental hidden scalar particles: \textit{Dakshina ($\chi$)} and \textit{Vama ($\xi$)} in nature.  In the lepton sector, there are only two fundamental leptons: electron ($e$) and electron type neutrino ($\nu_e$).  The $\nu_e$ is a left handed Dirac particle and its right handed counter part does not exist.  For the $e$, both the left handed and the right handed counterparts exist.
   \item The fundamental particles are associated with a hidden charge ($Q_h$),  a weak isospin charge ($T^3$) and a hidden isospin charge ($T^3_h$).  A bound state is formed such that its net hidden charge is zero and its net weak isospin and net hidden isospin charges are both $\pm 1/2$.
   \item There are two kinds of hidden gauge interactions: \textit{Shupti (Sh)} and \textit{Sushupti (Ssh)} in addition to EM and weak interactions.
   \item The fundamental particles $e,  \chi$ and $ \xi$ are massive while $\nu_e$ is massless.  All the massive fundamental particles get their masses through the Higgs mechanism.
   \item The masses of the fundamental hidden scalars are given in the following order:
   \begin{equation}
 M_\chi \lesssim M_\xi \sim 10^2 GeV
   \end{equation}
   \item The hidden scalars $\chi$ and $\xi$ are present as a uniform background in space.
   \end{enumerate}
   
     (The words \textit{Dakshina},  \textit{Vama},  \textit{Shupti} and \textit{Sushupti} are taken from Sanskrit which mean right, left, state of sleep and state of deep sleep respectively).

In this model,  only the first generation of the leptons are fundamental particles. The $e$ and the $\nu_e$ are  spin doublets ($2_s$),  while $\chi$ and $\xi$ are spin singlet($1_s$) states.  Hence,  the bound state of a fundamental fermion and a hidden scalar is a spin doublet state, $2_s\otimes 1_s = 2_s$. Thus, we obtain the  generations of the leptons as shown below.

\begin{eqnarray}
\begin{pmatrix}
e\\
\nu_e
\end{pmatrix}, \quad \begin{pmatrix}
e\ \chi\\
\nu_e\ \chi
\end{pmatrix}, \quad \begin{pmatrix}
e\ \xi\\
\nu_e\ \xi
\end{pmatrix}.
\end{eqnarray}

We identify the bound states in the following way,
\begin{eqnarray}
\mu = (e\chi), \quad\nu_{\mu}= (\nu_e\,\chi),\quad \tau = (e\,\xi), \quad\nu_{\tau}= (\nu_e\,\xi).
\end{eqnarray}

Thus, we see that the $e$ and the $\mu$ (or $\tau$) cannot be treated on an equal footing and the same holds for $\nu_{e}$ and $\nu_{\mu}$ (or $\nu_\tau$), too.  In addition,  we see that all three generations now share the same lepton number, $L_e$ and the necessity of individual lepton number vanishes. These features go against the primary axioms of the SM.

The  particles $\mu$ and $\tau$ are hidden-charge neutral and have EM charges as $-1$, whereas the composite neutrinos ($\nu_\mu$ and $\nu_\tau$)  are neutral in terms of both hidden and EM charges.  Hence,  the fundamental fermions and the hidden scalars must carry equal and opposite hidden charges.  We assign the hidden charges to the particles as shown below.
\begin{eqnarray}
Q_{h}(e,\nu_e)= -1,\quad Q_h(\chi,\xi)=+1.
\end{eqnarray} 

The $Shupti$ interaction is responsible for the formation of the bound states, whereas the $Sushupti$ is experienced only by the hidden scalars $\chi$ and $\xi$. We consider that $Sushupti$ converts $\chi$ to $\xi$ and vice versa.  The gauge groups associated with the $Shupti$ and the $Sushupti$ interactions are $U(1)_{h}$ and $SU(2)_h$ respectively. The $U(1)_h$ is generated by $Q_{h}$, whereas the hidden isospin charges, $T_{h}^{i}=\tau^{i}/2$ generate $SU(2)_h$,  where $\tau^i$'s, $i=1,2,3$ are the Pauli spin matrices. So, in the light of hidden interactions,  the SM gauge group is extended to $SU(2)_L\otimes U(1)_{Y}\otimes U(1)_h\otimes SU(2)_h$. The charges associated with the particles are described in Table.\,(\ref{table1})

We choose the left-handed fermions,
\begin{eqnarray}
L=\begin{pmatrix}
\nu_{eL}\\
e_L
\end{pmatrix}
\end{eqnarray}
as a doublet under $SU(2)_L$ and a singlet under $SU(2)_h$. It transforms under $U(1)_{Y}$ and $U(1)_{h}$. The right handed electron,  $e_R$ is chosen as a singlet under both $SU(2)_L$ and $SU(2)_h$ and it transforms under both $U(1)_{Y}$ and $U(1)_{h}$.
\begin{eqnarray}
L &\overset{U(1)_Y}{\longrightarrow}& e^{-i\,\frac{Y}{2}\alpha(x)}\,L = e^{\frac{+i}{2}\alpha(x)}\,L,\nonumber\\
L &\overset{U(1)_h}{\longrightarrow}& e^{-i\,Q_h\beta(x)}\,L = e^{+i\beta(x)}\,L,\nonumber\\
L &\overset{SU(2)_L}{\longrightarrow}& e^{-i\,\frac{\tau^i}{2}\theta^{i}(x)}\,L,\nonumber\\
R &\overset{U(1)_Y}{\longrightarrow}& e^{-i\,\frac{Y}{2}\alpha(x)}\,R = e^{+i\alpha(x)}\,R,\nonumber\\
R &\overset{U(1)_h}{\longrightarrow}& e^{-i\,Q_h\beta(x)}\,R= e^{+i\beta(x)}\,R,\nonumber\\
L &\overset{SU(2)_h}{\longrightarrow}& L'=L,\quad R \overset{SU(2)_L}{\longrightarrow} R'=R,\nonumber\\
R &\overset{SU(2)_h}{\longrightarrow}& R'=R.\nonumber
\end{eqnarray}

The hidden scalar fields $\chi$ and $\xi$ form a doublet under $SU(2)_h$,
\begin{eqnarray}
\phi_h= \begin{pmatrix}
\chi\\
\xi
\end{pmatrix},
\end{eqnarray}
and also transform under $U(1)_h$.  They transform as a singlet under both $U(1)_{Y}$ and $SU(2)_L$.   
\begin{eqnarray}
\phi_h &\overset{U(1)_h}{\longrightarrow}& e^{-i\,Q_h\beta(x)}\,\phi_h = e^{-i\beta(x)}\,\phi_h,\nonumber\\
\phi_h &\overset{SU(2)_h}{\longrightarrow}& e^{-i\,\frac{\tau^i}{2}\kappa^{i}(x)}\,\phi_h,\nonumber\\
\phi_h &\overset{U(1)_Y}{\longrightarrow}& \phi'_h=\phi_h,\quad \phi_h \overset{SU(2)_L}{\longrightarrow} \phi'_h=\phi_h.\nonumber
\end{eqnarray}

The complex scalar fields,
\begin{eqnarray}
\Phi=\begin{pmatrix}
\phi^+\\
\phi^0
\end{pmatrix},
\end{eqnarray}
form a doublet under $SU(2)_L$.  They transform under $U(1)_Y$,  but behave as a singlet under both $U(1)_h$ and $SU(2)_h$. 
\begin{eqnarray}
\Phi &\overset{U(1)_Y}{\longrightarrow}& e^{-i\,\frac{Y}{2}\alpha(x)}\,\Phi = e^{-\frac{i}{2}\alpha(x)}\,\Phi,\nonumber\\
\Phi &\overset{SU(2)_L}{\longrightarrow}& e^{-i\,\frac{\tau^i}{2}\theta^{i}(x)}\,\Phi,\nonumber\\
\Phi &\overset{U(1)_h}{\longrightarrow}& \Phi'=\Phi,\quad \Phi \overset{SU(2)_h}{\longrightarrow} \Phi'=\Phi.\nonumber
\end{eqnarray}

\begin{table}
\begin{center}
\begin{tabular}{|c||c|c|c|c|c|}
\hline
{}& $Q_{em}$ & $Y$& $T^3$&$Q_h$ & $T^3_h$\\
\hline
\hline
$\nu_{eL}$ & $0$ & $-1$ & $+\frac{1}{2}$ & $-1$ & $0$\\
\hline
$e_{L}$ & $-1$ & $-1$ & $-\frac{1}{2}$ & $-1$ & $0$\\
\hline
$e_{R}$ & $-1$ & $-2$ & $0$ & $-1$ & $0$\\
\hline
$\chi$ & $0$ & $0$ & $0$ & $+1$ & $+\frac{1}{2}$\\
\hline
$\xi$ & $0$ & $0$ & $0$ & $+1$ & $-\frac{1}{2}$\\
\hline
$\phi^+$ & $+1$ & $+1$ & $+\frac{1}{2}$ & $0$ & $0$\\
\hline
$\phi^0$ & $0$ & $+1$ & $-\frac{1}{2}$ & $0$ & $0$\\
\hline
\end{tabular}
\end{center}
\caption{\label{table1} \footnotesize The charges associated with different fields for $SU(2)_L\otimes U(1)_{Y}\otimes U(1)_h\otimes SU(2)_h$ theory are shown.}
\end{table}

The above transformation rules say that both $L$ and $R$ participate in both the EW and hidden interactions, while $\Phi$ and $\phi_h$ participate in the EW interaction and the hidden interaction respectively. The covariant derivative of this theory is shown below,

\begin{eqnarray}
D_\mu=\partial_{\mu} -i\, g \frac{\tau^i}{2}W_{\mu}^i - i\, g'\frac{Y}{2} B_{\mu}- i\, g_h \frac{\tau^i}{2}G_{\mu}^i-i\, g'_h Q_{h} C_{\mu},
\end{eqnarray}
where, $B_\mu$ and $W_{\mu}^i$ represent the gauge bosons associated with the EM and the weak interactions.  On the other hand,  $C_\mu$ and $G_{\mu}^i$ are related with the hidden interactions.
 
We present the Lagrangian ($\mathcal{L}$) describing the ``EW + Hidden'' forces under $SU(2)_L\otimes U(1)_Y\otimes SU(2)_h\otimes U(1)_h$ gauge symmetry as in the following,
\begin{eqnarray}
\mathcal{L}= \mathcal{L}_F + \mathcal{L}_H +\mathcal{L}_S +\mathcal{L}_G +\mathcal{L}_Y,
\end{eqnarray} 
where, 
\begin{eqnarray}
\mathcal{L}_F &=& i\,\bar{L}\gamma^\mu(\partial_\mu-i\,g \frac{\tau^i}{2}W_{\mu}^i+i\, g' B_\mu)L + i\,\bar{e}_R \gamma^{\mu} (\partial_\mu + i\,g'B_{\mu})e_R - g'_h(\bar{L}\gamma^\mu L +\bar{e}_R \gamma^\mu e_{R})C_\mu,\nonumber\\
\\
\mathcal{L}_h &=& \lbrace(\partial^\mu -i\,g_h \frac{\tau^i}{2}G^{\mu^{i}}-i\,g'_h C^{\mu})\phi_h \rbrace^{\dagger} \lbrace(\partial_\mu -i\,g_h \frac{\tau^i}{2}G_{\mu}^{i}-i\,g'_h C_{\mu})\phi_h \rbrace,\\
\mathcal{L}_S &=& \lbrace(\partial^\mu -i\,g \frac{\tau^i}{2}W^{\mu^{i}}-i\,g' B^{\mu})\Phi \rbrace^{\dagger} \lbrace(\partial_\mu -i\,g \frac{\tau^i}{2}W_{\mu}^{i}-i\,g' B_{\mu})\Phi \rbrace + \mu^2 (\Phi^\dagger\Phi)-\lambda(\Phi^\dagger \Phi)^2,\nonumber\\
\mathcal{L}_G &=&-\frac{1}{4}B_{\mu\nu}B^{\mu\nu}-\frac{1}{4}W^{i}_{\mu\nu}W^{{\mu\nu}^i}-\frac{1}{4}C_{\mu\nu}C^{\mu\nu}-\frac{1}{4}G^{i}_{\mu\nu}G^{{\mu\nu}^i},
\end{eqnarray}
where, $\mu^2>0, \lambda>0$. The $B_{\mu\nu}$, $W_{\mu\nu}^i$, $C_{\mu\nu}$, and $G_{\mu\nu}^i$, represent the field tensors for the $B_{\mu}$, $W_{\mu}^i$, $C_\mu$, and $G_{\mu}^i$ respectively. Here, the $\mathcal{L}_Y$ is the Yukawa term and this is presented below.
\begin{eqnarray}
\mathcal{L}_Y= -y_e (\bar{L}  \Phi e_R)-y_{\chi}^2 (\phi_h ^{\dagger} \Phi) (\Phi^{\dagger}\phi_h)-y_{\xi}^2 (\phi_h ^{\dagger} \tilde{\Phi}) (\tilde{\Phi}^{\dagger}\phi_h)+ H.C.
\end{eqnarray}

We can see that only the mass terms of the fundamental particles are present in the Yukawa term.

The $SU(2)_L$ scalar field doublet, $\Phi$ is parametrized in the following way,
\begin{equation}
\Phi=\frac{1}{\sqrt{2}}e^{-i\frac{\tau^i\rho^{i}(x)}{2v}}\begin{pmatrix}
0\\
v+H
\end{pmatrix},
\end{equation}
where,  $v$ is the vacuum expectation value, $v=\sqrt{\mu^2/\lambda}$,  $\rho^i(x)$ are the ``would be Goldstone bosons''. We take the unitary gauge which involves a gauge transformation by a $SU(2)_L$ element, $U(\rho^i)=e^{i\frac{\tau^i\rho^{i}(x)}{2v}}$. This prevents the Goldstone's boson to appear in the final  equation of motion and the $\Phi$ becomes real,
\begin{equation}
\Phi=\frac{1}{\sqrt{2}}\begin{pmatrix}
0\\
v+H
\end{pmatrix},
\end{equation}

where, $H$ is the Higgs field. The unitary gauge changes: $L\rightarrow L'$, $W_{\mu}^i\rightarrow W_{\mu}^{i'}$, $B_\mu$ and $e_R$ do not change.   (Note that in the subsequent discussion, in order to keep the expressions simple, we shall write the primed fields as the original ones). The hidden scalar field $\phi_h$  being an $SU(2)_L$ singlet, remains unaffected. Similarly, the unitary gauge has no effect on the hidden gauge fields $C_\mu$ and $G_{\mu}^i$.  The Higgs mechanism breaks the  $SU(2)_L\otimes U(1)_Y\otimes SU(2)_h\otimes U(1)_h$ gauge symmetry to $U(1)_{e.m}\otimes U(1)_h$.  All the hidden gauge bosons along with the physical photon field $A_{\mu}$ remain massless.  Our axiom, however, dictates that the hidden scalars, $\chi$ and $\xi$ acquire unequal masses. Therefore, the  $SU(2)_h$  symmetry is not preserved.  The fact that the carriers of the hidden forces are massless actually makes them cosmologically stable and they could be projected as viable candidates for DM. This is based on the cosmological stability condition for vector bosons ($m_V < 2m_e$) to be DM candidates\, \cite{ParticleDataGroup:2020ssz}.  We may split the Lagrangian into two parts: the first part is from the GWS framework, $\mathcal{L}_{GWS}$ which tells us about the dynamics of $e$, $\nu_e$, $H$ and how they interact via EM and weak interactions; and the second part which is $\mathcal{L}_{hidden}$ tells us about the hidden scalar particles $\chi$ and $\xi$, and how these particles along with $e$ and $\nu_e$ take part in the hidden interactions: $Shupti$ and $Shusupti$. It also tells us how $\chi$ and $\xi$ interact with $H$.  We present the Lagrangian as shown below,

\begin{eqnarray}
\mathcal{L}_{U(1)_{Q}\otimes U(1)_h} &=& \mathcal{L}_{GWS} + \mathcal{L}_{hidden}.
\end{eqnarray}

where,

\begin{eqnarray}
\mathcal{L}_{GWS} &=& i\,\bar{e}\gamma^{\mu}\partial_\mu e +i\,\bar{\nu}_{eL}\gamma^\mu \partial_\mu \nu_{eL} -\frac{y_{e} v}{\sqrt{2}} \bar{e}\,e\nonumber\\
&& + \frac{g}{\sqrt{2}}\left( \bar{\nu}_{eL}\,\gamma^{\mu}\,e_L\, W_{\mu}^+ +\bar{e}_{L}\,\gamma^{\mu}\,\nu_{eL}\, W_{\mu}^{-}\right)\nonumber\\
&& + \frac{g}{2}\,\left(\bar{\nu}_{eL}\,\gamma^\mu\,\nu_{eL}-\bar{e}_L\,\gamma^\mu\,e_L\right)\left( Z_\mu\cos\theta_W + A_\mu \sin\theta_W\right)\nonumber\\
&&- \frac{g'}{2}\,\left(\bar{\nu}_{eL}\,\gamma^\mu\,\nu_{eL}+\bar{e}_L\,\gamma^\mu\,e_L + 2 \bar{e}_R\,\gamma^\mu\,e_R \right)\left( A_\mu\cos\theta_W - Z_\mu \sin\theta_W\right)\nonumber\\
&&+\frac{g^2 v^2}{4}W_{\mu}^+ W^{\mu^-}+ \frac{v^2}{8}\,(g^2+g'^2)Z^{\mu} Z_{\mu}\nonumber\\
&&+\,\partial_\mu H\,\partial^\mu H - \mu^2\,H^2-\lambda\,v\,H^3-\frac{\lambda}{4}\,H^4\nonumber\\
&&-\frac{y_e}{\sqrt{2}}\,H \bar{e}e + \frac{g^2}{8}(H^2 + 2\,H v)\left(\frac{1}{\cos^2\theta_{W}}Z^{\mu} Z_{\mu}+ 2 W_{\mu}^+ W^{\mu^-}\right),
\end{eqnarray}
where, $W_\mu^{+}=\frac{1}{\sqrt{2}}\,(W_\mu^1-i\,W_\mu^2)$, $W^-_\mu=\frac{1}{\sqrt{2}}\,(W_\mu^1+i\,W_\mu^2)$ and, $\theta_{W}=\tan^{-1}(g'/g)$ and,

\begin{eqnarray}
\mathcal{L}_{hidden}&=& \partial ^\mu\,\chi^* \partial_\mu \chi- y_{\chi}^2 v^2\,\chi^*\chi +\partial ^\mu \xi^*\, \partial_\mu \xi- y_{\xi}^2v^2\,\xi^*\xi\nonumber\\
&& + \frac{g_h}{2}\, \lbrace i\, (\chi^*\, \partial^\mu \chi-\chi\,\partial^\mu \chi^*)- i\,(\xi^* \,\partial^\mu \xi - \xi\,\partial^\mu \xi^*)\rbrace\,G_{\mu}^3\nonumber \\
&& + \frac{g_{h}}{2}\, i\,(\xi^*\,\partial^\mu \chi-\chi\, \partial^\mu \xi^*)\,G^*_\mu + \frac{g_{h}}{2}\, i\,(\chi^*\,\partial^\mu \xi-\xi\,\partial^\mu \chi^*)\,G_\mu \nonumber\\
&& +\, g'_h\,\lbrace i\,(\chi^*\, \partial^\mu \chi-\chi\,\partial^\mu \chi^*) + i\,(\xi^*\, \partial^\mu \xi-\xi\,\partial^\mu \xi^*)-(\bar{e}\gamma^\mu e +\bar{\nu}_{eL} \gamma^{\mu}\nu_{eL})\rbrace\,C_{\mu} \nonumber\\
&& +\left( \frac{g_h}{2}\,(\chi^*\,G^{\mu^3}+G^{\mu^*}\,\xi^*)+g'_h \chi^*\,C^\mu \right)\left( \frac{g_h}{2}\,(\chi\,G^{^3}_\mu + G_\mu\,\xi)+g'_h \chi\,C_\mu \right)\nonumber\\
&&+\left( \frac{g_h}{2}\,(\xi^*\,G^{\mu^3}-G^\mu\,\chi^*)-g'_h \xi^*\,C^\mu \right)\left( \frac{g_h}{2}\,(\xi\,G^{^3}_\mu + G^*_\mu\,\chi)-g'_h \xi\,C_\mu \right)\nonumber\\
&&-(H^2+2\,v\,H)(y_{\chi}^2\,\chi^*\chi +y_{\xi}^2\,\xi^*\xi),
\end{eqnarray}
where, $G_\mu=(G_\mu^1-i\,G_\mu^2)$, and $G^*_\mu=(G_\mu^1+i\,G_\mu^2)$. 

The $Shupti$ is associated with the $C_\mu$ vector field which will be referred to as ``hidden photon''($\gamma_h$) in subsequent discussion. Similarly, $Sushupti$ is mediated by the hidden vector bosons, $G_{\mu}$, $G_{\mu}^*$ and $G_\mu^3$ and these are represented as $G$, $G^*$ and $G^3$ respectively. 
\vspace{2mm}

We discuss some important points of our model below:

\begin{enumerate}

\item[(a)]We see that the hidden photon field ($C_\mu$) like the physical photon field ($A_\mu$) is massless, but unlike the latter, the former may interact with the electrically neutral particles. We get the the following primary vertices for the $Shupti$ interaction (see Fig.\,(\ref{shupti})),
\begin{eqnarray}
e^{-}&\rightarrow & e^- +\gamma_h,\nonumber\\
\nu_{e}&\rightarrow &\nu_{e} +\gamma_h,\\
\chi &\rightarrow & \chi +\gamma_h,\\
\xi &\rightarrow & \xi + \gamma_h.
\end{eqnarray}

\begin{figure}
\begin{center}
\includegraphics[scale=0.2]{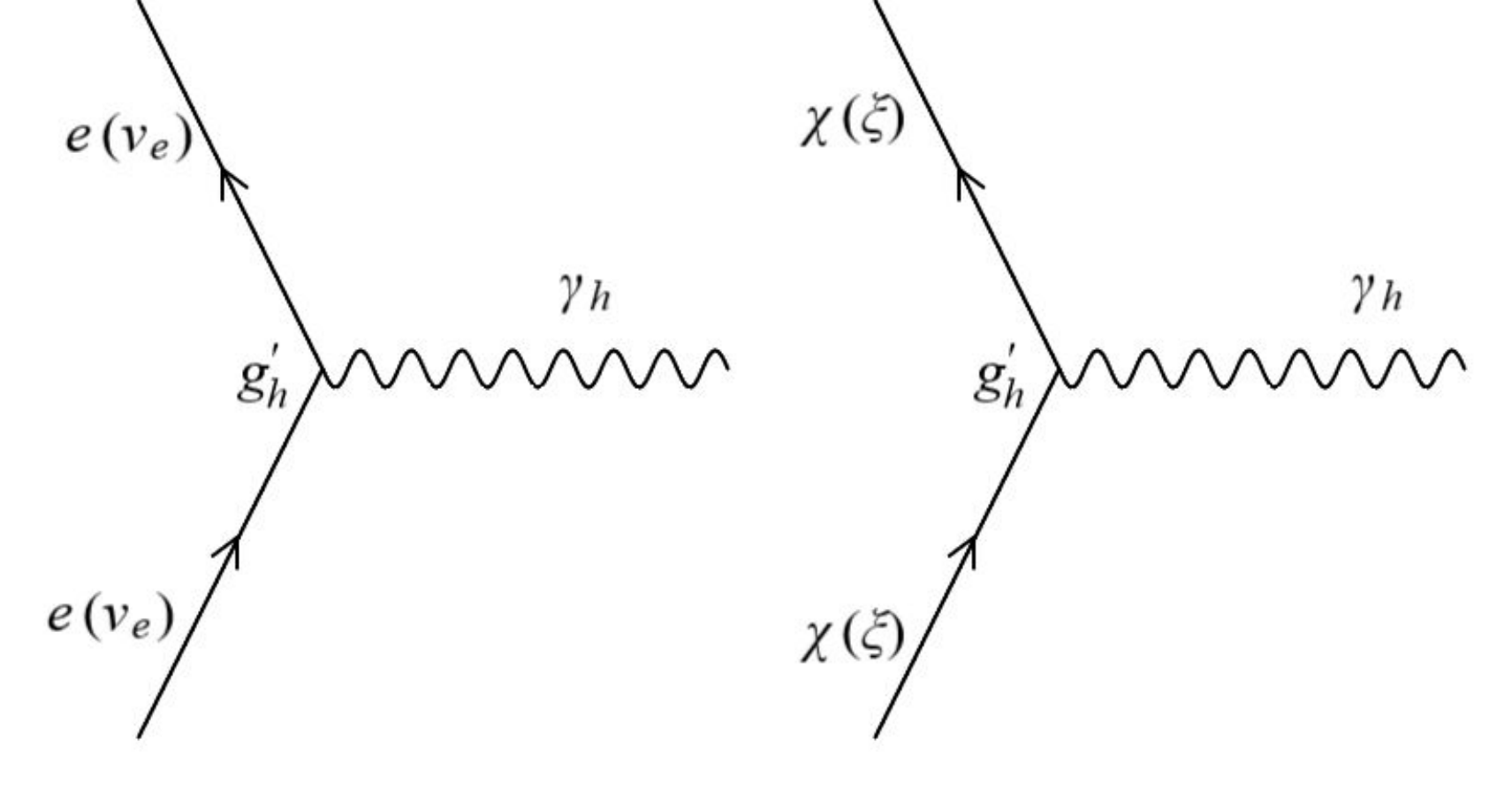}
\caption{\label{shupti} The primary vertices for $Shupti$ interaction.}
\end{center}
\end{figure}

\item[(b)] Similar to the $Sh$ interaction, the $Ssh$ force is associated with `hidden isospin raising' (HIRC), `hidden isospin lowering' (HILC) and `hidden neutral isospin' (HNIC) currents. The corresponding primary vertices are shown below.
\begin{eqnarray}
\xi &\rightarrow & \chi+ G, \hspace{0.2 cm} (HIRC)\\
\chi &\rightarrow & \xi+ G^*, \hspace{0.2 cm} (HILC)\\
\xi\,(\chi) &\rightarrow & \xi\,(\chi)+ G^3. \hspace{0.2 cm} (HINC)
\end{eqnarray}

The Feynman diagrams are depicted in Fig.\,(\ref{sushupti})

\begin{figure}
\begin{center}
\includegraphics[scale=0.2]{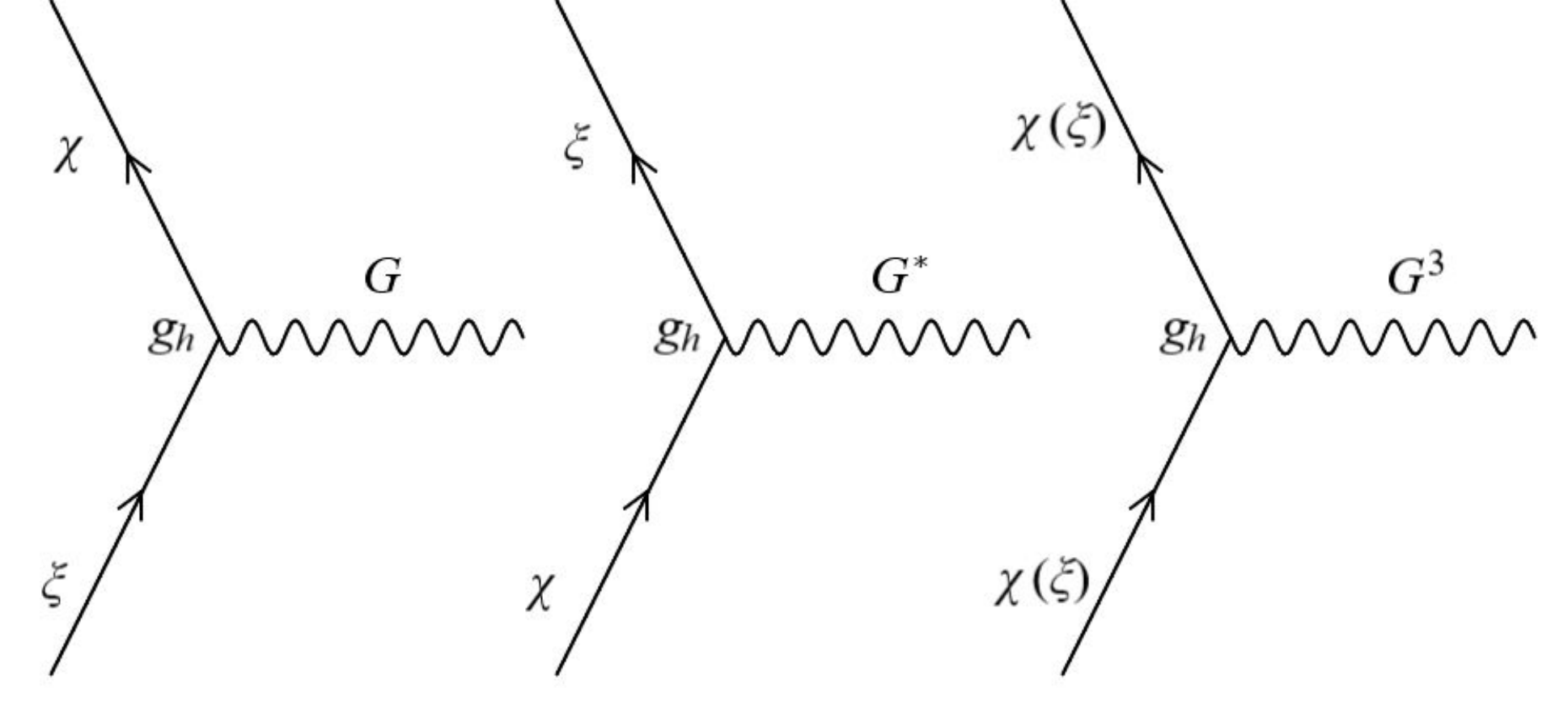}
\caption{\label{sushupti} The primary vertices for $Sushupti$ interaction.}
\end{center}
\end{figure}

\item[(c)] In addition to the bound states like $\mu$ and $\tau$,  we might wonder whether we could find hidden bound states like $\chi\bar{\chi}$, $\xi\bar{\xi}$, $\chi\bar{\xi}$ and $\bar{\chi}\xi$ or dileptonic states like $e\bar{\nu}_e$, $\bar{e}\nu_e$ and $\nu_e\bar{\nu}_e$ in nature or not.  It is seen that although we have $Q_h=0$ for these states,  they are discarded by the restrictions $T^3$(bound state)=$T^3_h$ (bound state)=$\pm 1/2$. So, they are not found in nature.  However, there might be a follow up question regarding the non-observation of composite states with more than two constituents,  e.g.,  $e\chi e\bar{e}$,  $e\xi\xi\bar{\xi}$ etc.  The above restrictions cannot disallow these states.  The explanation for their non existence might require the presence of a new force,  which is discussed in \ref{subsection 3C}.

\end{enumerate}

\section{\label{Section 3}Features of the extended GWS Model} 
In this section,  we reinterpret two phenomena, viz., the decay of the charged leptons under the weak interaction and the neutrino oscillation in the light of our model.  We also touch upon the question of flavour violation in the charged lepton sector.  Finally,  we discuss the mass generation of the composite leptons and determine the mass scales of the contributions from the hidden interactions to the same.

\subsection{\label{subsection 3A} {Muon Decay}}
 In the light of the proposed model, it is possible to explain the decay processes of leptons. To exemplify,  we choose the $\mu$ decay. We know that the SM depicts muon decay in the following way, 
\begin{eqnarray}
\mu\rightarrow e + \nu_\mu + \bar{\nu}_e .
\end{eqnarray}
This process is explained in the light of the composite nature of $\mu^-$ as in the following: the $e$ present in the $\mu$, decays to $\nu_e$ and $W^-$; the $\chi$ sticks to $\nu_e$ and forms $\nu_\mu$ and the $W^-$ decays to $e$ and $\bar{\nu}_e$ (See Fig.\,(\ref{muondecay})). The process is depicted below,
\begin{eqnarray}
\mu\,(e\,\chi)\rightarrow \nu_{\mu}\,(\nu_e\,\chi) +\, W^-;\quad W^-\rightarrow e + \bar{\nu}_e\nonumber
\end{eqnarray}

\begin{figure}
\begin{center}
\includegraphics[scale=0.2]{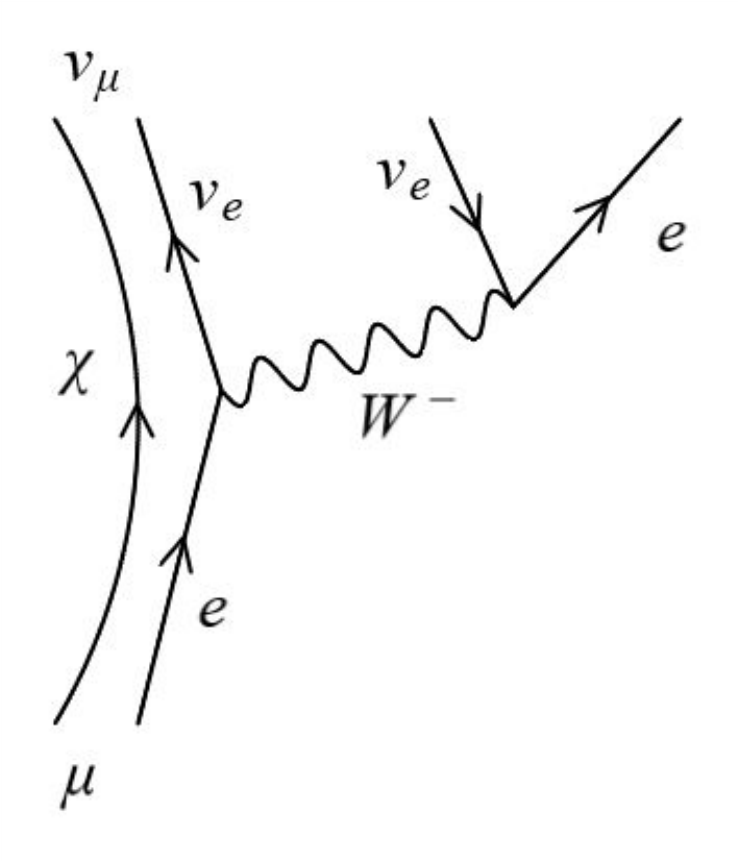}
\end{center}
\caption{\label{muondecay} The muon decay process is depicted in the light of the proposed model. In the shown diagram, time flows in the upward direction. The muon is a bound state of $e$ and $\chi$. The electron decays to $\nu_{e}$ via emission of $W^-$. The $\nu_e$ and $\chi$ bind to form $\nu_{\mu}$.  The $W^{-}$ decays to $e$ and $\bar{\nu}_e$.}
\end{figure}

\subsection{\label{subsection 3B}{Neutrino Oscillation}}
Our model can explain neutrino oscillation as a dynamical phenomenon. It says that the neutrino oscillation occurs as a result of the interaction of the two hidden forces with the $\nu_e$ and the hidden scalars.  As the hidden scalars are present as a uniform background,  the $\nu_e$,  while it travels,  may get bound to them and subsequently,  get split up.  So we get the following oscillations,
\begin{eqnarray}
\nu_e + \chi \longleftrightarrow \nu_\mu,\\
\nu_e + \xi \longleftrightarrow \nu_\tau
\end{eqnarray}
To an observer, $\nu_e$ simply changes its flavour to $\nu_{\mu}$ (or $\nu_{\tau}$) and similarly, when this bound state is broken, the $\chi$ (or $\xi$) is released and one observes the same only as a flavour conversion from $\nu_{\mu}$ (or $\nu_\tau$) to $\nu_e$. 

And the conversion of $\nu_\mu$ to $\nu_\tau$ and vice versa can be understood in terms of the $Sushupti$ interactions which change $\chi$ to $\xi$ and vice versa. 
 
 \begin{eqnarray}
\nu_{\mu}&\rightarrow & \nu_{\tau} +G^*,\quad G^* \rightarrow  \chi + \bar{\xi},\\
\nu_{\tau}&\rightarrow & \nu_{\mu} + G ,\quad G \rightarrow  \bar{\chi} + \xi
\end{eqnarray}

As the hidden particles are released, the detectors cannot detect those particles. So to an observer, this would simply look like the flavour conversion of $\nu_\mu$ to $\nu_\tau$ and vice versa.

\begin{figure} 
\begin{center}
\includegraphics[scale=0.2]{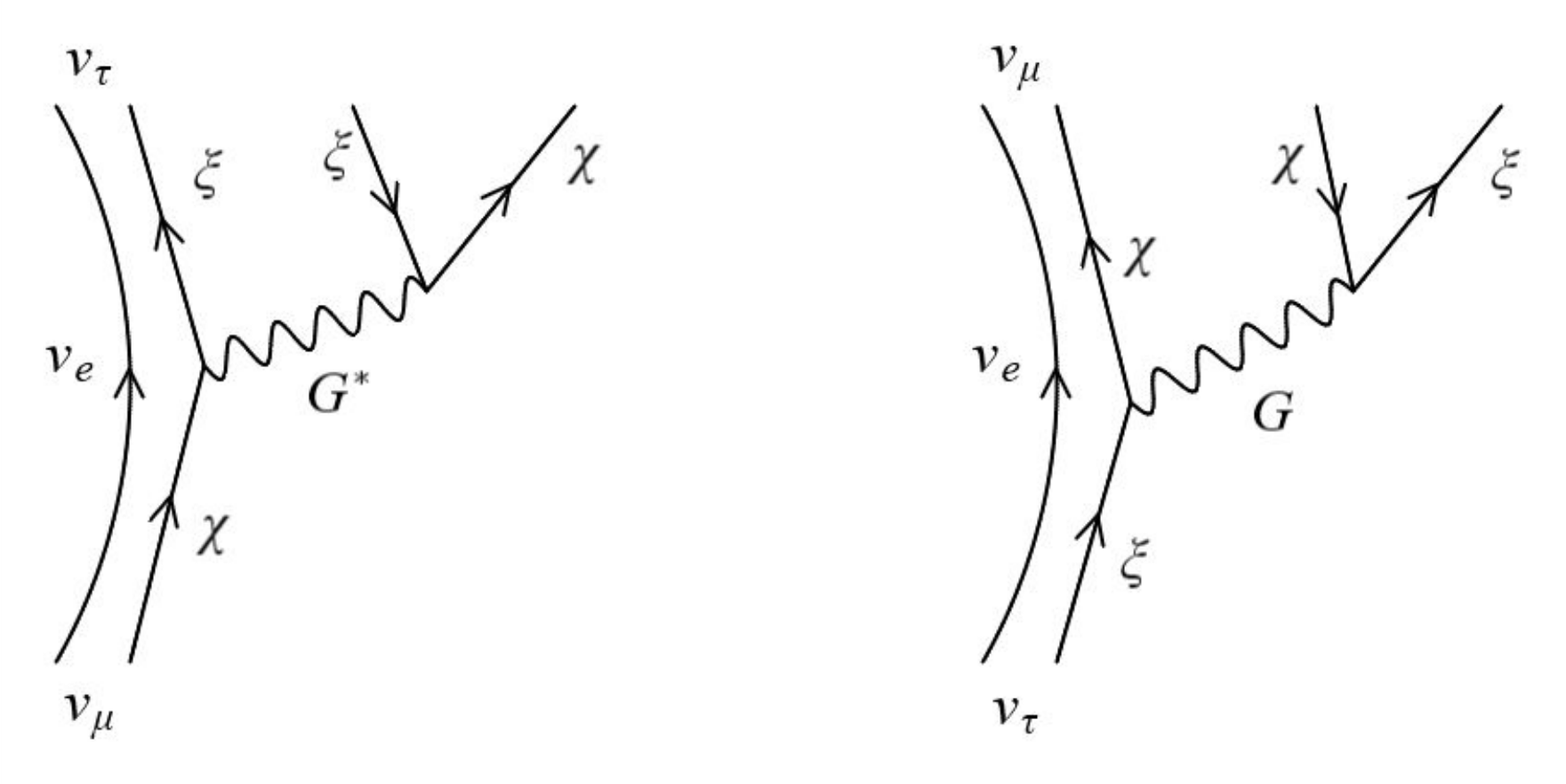}
\caption{\label{neutrinooscillation} The neutrino flavour oscillation between $\nu_{\mu}$ and $\nu_{\tau}$ is described in the light of the proposed model. In the above Feynman diagrams, the time flows in the upward direction. The $\nu_{\mu}$ is composed of $\nu_e$ and $\chi$. In the left diagram, the $\chi$ decays to $\xi$ via emission of a $G^*$. The $\xi$ binds itself to the $\nu_e$ and forms a $\nu_{\mu}$. The $G^*$ decays to $\chi$ and $\bar{\xi}$. The hidden scalar particles are unobservable and hence it appears that $\nu_{\mu}$ is converted to $\nu_{\tau}$. Similarly, in the right diagram,$\nu_{\tau}$ is converted to $\nu_{\mu}$ via the emission of $G$.}
\end{center}
\end{figure}

The relevant Feynman diagrams are given in the Fig.\,(\ref{neutrinooscillation})

A few remarks about the above discussion are due here.  First,  it is seen that our model treats neutrino oscillation in terms of the two hidden interactions in two different ways.  Oscillations that involve $\nu_e$ directly can be described by the $Shupti$ force,  while oscillations not involving $\nu_e$ directly can be described by the $Sushupti$ force.  The $\nu_\mu \longleftrightarrow \nu_\tau$ oscillation can be described as a combination of neutrino flavour conversion and invisible decay\, \cite{Bahcall:1972my,Acker:1991ej,Lindner:2001fx,Barger:1998xk} as the hidden scalars are released in both cases.   Here the oscillation time might be determined from the lifetime of the hidden scalars under the $Sushupti$ decay. The oscillation $\nu_e \longleftrightarrow \nu_\mu(\nu_\tau)$ is a little complicated as it involves formation and breaking of a bound state. The hidden scalars present as a uniform background will also play a role here.  Both cases are in contrast to the PMNS model where the quantum interference of the mass eigenstates is responsible for neutrino oscillation.   

\subsection{\label{subsection 3C}{Flavour Violation in the Charged lepton Sector}}
 Conservation of lepton flavour (LF) or lepton family number is a consequence of the $U(1)_e \otimes U(1)_\mu \otimes U(1)_\tau$ symmetry of the SM,  which is an accidental global symmetry\,\cite{Ardu:2022sbt,deGouvea:2013zba}.  An accidental symmetry is not a fundamental symmetry of nature that has to be conserved at all energy scales,  but it is a symmetry that happens to be conserved at a given energy scale where relevant symmetry breaking operators are not allowed by theory.  However,  those symmetry breaking operators might arise at higher energy scales beyond the jurisdiction of the concerned low energy theory.  In that case,  those operators are said to represent new physics and as such,  accidental symmetries can be violated in the realm of new physics.  
 
 The invariance of the SM Lagrangian under the said global symmetry is based on the  assumption that neutrinos are massless.  But it is known that neutrinos have masses and their mass generation mechanism must involve new physics beyond the SM (BSM).  This new physics should break the LF symmetry in the neutrino sector,  which is evident in the phenomenon of $\nu$ oscillation.  But the existence of such new physics will also have impact upon the charged lepton sector as neutrinos are related to their charged lepton partners by $SU(2)_L$ symmetry.  Hence,  it is not illogical to expect  flavour violation in the charged lepton sector too.  However, this argument fails if we take into account the expected upper bound on the branching ratio (BR) of the charged lepton flavour violating (CLFV) process $\mu \longrightarrow e+\gamma$ based on neutrino as a source of lepton flavour violation. The value of the upper bound is too small, of the order of $10^{-54}$\,\cite{Workman:2022ynf},  to be experimentally detected.  This result shows that as far as CLFV is concerned,  LFV in neutrino sector might not play much role.  In addition, till date, there is no experimental evidence for CLFV.  For example, so far, the best experimental upper bounds on BR for CLFV processes like $\mu \longrightarrow e\gamma$ and $\mu \longrightarrow 3e$ are respectively $10^{-13}$ and $10^{-12}$\,\cite{Workman:2022ynf},  which are also rather small to be conclusive about the presence of CLFV. Now, we could definitely argue that with the increasing sensitivities of the experiments, improved upper bounds can be found, but there is also one possibility that perhaps nature disfavours CLFV.
 
 Regarding why CLFV might not be observed in nature,  we would like to follow the Gell-Mann's totalitarian principle\,\cite{Gell-Mann:1956iqa},  which states that if a process is not forbidden by a fundamental principle,  it is bound to occur.  So the apparent non observation of CLFV processes might point to some new fundamental hidden mechanism underlying it which is not present in the neutrino sector.  
  
  Our model,  motivated by the fact that LF symmetry is not a fundamental symmetry of nature,  does not consider differences in the species of the leptons based on lepton family numbers or flavours,  but based on their compositions.  In our model,  there is only one kind of lepton number (the electron number),  which is always conserved.  This is also expected from our axiom that the $\nu$ is of Dirac nature.  The gauge forces introduced allow conversion from one species into another (see \ref{subsection 3B}).  However,  the same forces can induce similar conversions in the charged lepton sector, too.   So,  to account for the non-observation of such processes,  at least in low energies,  we postulate the possibility of a new force that could exist between the $e$ and the hidden scalar particles.  This new interaction should,  however,  not be experienced by neutrinos.  Hence,  this new force should depend on quantities that the $e$ possesses,  but the $\nu_e$ does not.  The quantities that differentiate an electron from a neutrino are EM charge and mass.  But the hidden scalars do not possess EM charges.  So the new force that distinguishes the $\nu_e$ from its charged lepton partner should depend on mass.  We see that this force does not obey $SU(2)_L$ symmetry as it treats $e_L$ and $(\nu_e)_L$ on different footings.  Further, we propound that this new force is also responsible for the non-observation of many particle (more than two) composite states.  A detailed study on this force is beyond the scope of this paper,  but we posit the minimum requirements that this force must obey as given below:
\begin{itemize}
\item The force is a gravity motivated force,  i.e,  it is proportional to the product of the constituent masses  and it exists when both the constituents are massive.
\item Unlike gravity,  this force is a repulsive force.  The repulsive barrier that needs to be overcome to form a two particle bound state is referred to as the charged lepton fusion (CLF) scale.  Further,  the repulsive power of the force in a bound state towards a third particle is  far greater than the same in an individual fundamental particle.
\item The force prevents hidden isospin change of the constituents, although it allows weak isospin change. The relevant selection rule would be $\Delta I_h^3 \neq \pm1$.
\item Once, a bound state is formed, it does not allow it to split up, i.e., the force strictly excludes other possibilities than $Q^{total}_h=0$ for a state in which it is present.
\end{itemize}

The first assertion rules out the participation of the composite neutrinos in this interaction, because in our model,  the mass of the fundamental neutrino,  $\nu_e$ is zero.  This means that the conversion of one $\nu$ species into another is allowed in the light of the new force.  The second assertion is necessary because it prevents the electron and the hidden scalars from being bound to each other at energies below the CLF scale, i.e., effectively discarding $e \longrightarrow \mu(\tau)$ transitions .  The CLF scale is identical to the CLFV scale.  The fact that the repulsive power of the force increases strongly once a bound state is formed, prevents further conglomeration that could have led to many particle composite states.  From the third assertion,  it is seen that $\chi$ can not change to $\xi$ and vice versa via a change of hidden isospin,  which rules out $\mu \longleftrightarrow \tau$ flavour conversions. The fourth assertion prevents the split up of a composite charged lepton because free fundamental particles have non zero hidden charges.   This might be the reason why processes like $\mu(\tau) \longrightarrow e + \gamma$ in the charged lepton sector are not seen.

\subsection{Mass of the Composite Particles}
In our model,  all the composite particles get masses from the intrinsic masses of the constituents and from the forces binding them.  Assuming that the new force discussed above exists in the composite charged leptons,  we write the expressions for the composite particle masses in the following way:
\begin{eqnarray}
M_\mu &=& M_e + M_\chi + C(Shupti) + D_1,\\
M_{\nu_{\mu}} &=& M_\chi + C(Shupti),\\
M_\tau &=& M_e + M_\xi + C(Shupti) + D_2,\\
M_{\nu_{\tau}} &=& M_\xi + C(Shupti),
\end{eqnarray}

where $C(Shupti)$ is the contribution from the $Shupti$ force and it is equal in all the bound states because it depends on the hidden charge of the constituents. $D_1$ and $D_2$ are the contributions from the new force and it is not equal in $\mu$ and $\tau$ because we have assumed $M_\chi \neq M_\xi$.

By solving the above equations,  we get the following results:
\begin{eqnarray}
 D_1 &=& M_\mu - (M_{\nu_\mu} + M_e) \sim  M_\mu \sim 10^2 MeV\\
 D_2 &=& M_\tau - (M_{\nu_\tau} + M_e) \sim M_\tau \sim 10^3 MeV\\
 M_\xi -M_\chi &=& M_{\nu_\tau}-M_{\nu_\mu}\\
 C &= &M_{\nu_\mu} - M_\chi \sim -M_\chi
\end{eqnarray}

So,  the contribution from the new force is of the order of the mass of the muon in case of the muon and the mass of the tauon in case of tauon.  Also, the difference in the masses of the hidden scalars is very small,  which is of the order of a fraction of an $eV$.  Again,  the contribution to the bound state masses from the $Shupti$ interaction is $\sim 10^2 GeV$.  It is also possible that the hidden scalars could be candidates for the scalar WIMP DM given their mass scales and their non-participation in the SM interactions\, \cite{Birkedal-Hansen:2003dym,Roszkowski:2017nbc}.

There are a few important points in this section that deserve mention here.  Regarding the new force that could exist within the composite charged leptons,  the repulsive character of the force is noteworthy.  It is  contrasting to the gravity which is always attractive in nature.  It is also interesting to note that the new force, despite being repulsive in nature, allows the bound states of  $\mu$ and $\tau$ to form beyond the CLF scale,  but prevents the same from breaking.  An important observation can be made from the muon decay: although this new force prevents splitting up of a composed charged lepton into its components,  it, however, allows weak decay to take place and the force also vanishes in the final products as $\nu_e$ is massless.  A thorough analysis of this observation could provide a vital platform to our quest to know more about this force. There are also some important questions to be answered such as whether such a force can be explained in terms of a gauge interaction\,\cite{Narain:2013eea,Deser:1974cy}.  This needs a detailed investigation of the renormalisability issue at high energies.  Again,  it is important to test the predictions of this force against the experimental constraints on flavour violating processes\,\cite{Baldini:2021kfb}.  For that, the knowledge of particle decay under this force would be important.  In case,  it turned out not to be  describable in terms of a gauge interaction,  the expression of the decay rates or the approach to finding it might be quite different.  This is another important point that needs to be explored.  Additionally,  the fact that this force is also responsible for the non-observation of the $N$ particle composed states, where $N > 2$ is interesting because it explains why many particle bound states like $e\chi e\bar{e}$,  $e\chi\chi\bar{\chi}$ etc. , are not found in nature at the present energy scale. This is again based on the repulsive nature of the force and the fact that this nature becomes more prominent in a bound state toward a third particle.  Formation of such many particle composite states would be possible at a high energy scale which would be sufficient to overcome the increasing repulsive force. A dedicated study,  in this regard,  is expected to reveal many new insights.  Besides,  the contributions from the hidden forces need to be derived from the theories of the forces themselves. The work is currently in progress.  We expect to discuss them in more details in our future works.  Further,  there might be questions regarding the mass scales of the hidden particles and the contributions from the hidden interactions to the composite particle masses.  These contributions and the masses of the constituents are within the reach of the current collision energy of the LHC, too\, \cite{Buarque:2021dji}.  So,  one could ask why the substructures are not found yet.  One possible reason is that the hidden particles do not take part in SM interactions.  As to the question of the existence of the hidden forces, we have shown that only the $e$ and the $\nu_e$ are the SM particles that participate in $Shupti$ interaction.  Existence of such a force can be tested indirectly via the impact due to the hidden charges ($Q_h$) on experimentally measured quantities.   Apart from this,  there are also technical challenges for detection of these particles\, \cite{Kane:2008gb,Schmidt:2016jra}.

\section{\label{Section 4}Summary and Discussion}

The present model,  unlike most of the other composite models,  does not insist on all the leptons having substructures nor does it demand compositeness of the gauge bosons, as well.  It posits that the $e$ and the $\nu_e$ are fundamental particles and the rest are composite particles.  The model,  however,  allows the hidden matter fields: $\chi$ and $\xi$ to reside within the framework of the SM.  The hidden matters do not interact via EW interactions,  rather they experience two new gauge interactions: $Shupti$ and $Sushupti$.   The gauge group that describes the EW and the hidden interactions is $SU(2)_L\otimes U(1)_Y\otimes SU(2)_h\otimes U(1)_h$ and the Higgs mechanism breaks the symmetry to $U(1)_Q\otimes U(1)_h$.  Through the Higgs mechanism,  all the fundamental particles except the $\nu_e$ obtain masses,  while $\nu_e$ remains massless. The model attributes the masses of the composite particles to their compositeness,  which includes the masses of the constituent fundamental particles and the contributions from the hidden forces.  It does not require the existence of the right handed Dirac neutrino or Majorana neutrino to explain neutrino masses.  The model can also reinterpret a few phenomena like the neutrino oscillation and the weak decay of charged leptons in the framework of the hidden interactions.  In addition,  to explain the absence of flavour violation in the charged lepton sector and why many particle bound states are not observed,  it is found that the existence of a new force between the electron and the hidden scalars is necessary.  The force,  as understood from the present context,  is repulsive, mass dependent and violates the $SU(2)_L$ symmetry.  A detailed picture of this force is beyond the scope of this paper. 

 An important feature of our model is that it does not treat the leptons and the quarks on the same footing.  It does not say anything about compositeness of the quarks.  In the light of quark-lepton symmetry,  as emphasised by the SM,  one may extend the present discussion to the quark sector as well,  as is done by many composite models.  But we see that the presence of the neutrinos within the lepton sector somehow makes the latter a little special and there lies the motivation to treat the leptons differently from the quarks at a more fundamental level. 

 The guiding philosophy of our model comes from the centuries old experience of mankind with nature,  which reminds us that the idea of a fundamental particle is not conclusive.  Sometimes,  nature gives us hints in the form of peculiarities of particles that cannot be explained by the existing theories. Our paper tries to address some of those issues by introducing a layer of compositeness,  while keeping the basic framework of the GWS model intact.

\section*{Acknowledgement}
The authors extend their gratitude to Prof. N. Nimai Singh,  Manipur University for a valuable discussion on neutrino oscillation.


\end{document}